\documentclass[conference]{IEEEtran}
\IEEEoverridecommandlockouts
\usepackage{cite}
\usepackage{amsmath,amssymb,amsfonts}
\usepackage{algorithm}
\usepackage{algorithmic}
\usepackage{graphicx}
\usepackage{textcomp}
\usepackage{xcolor}
\usepackage[caption=false, font=footnotesize]{subfig}
\def\BibTeX{{\rm B\kern-.05em{\sc i\kern-.025em b}\kern-.08em
    T\kern-.1667em\lower.7ex\hbox{E}\kern-.125emX}}
\usepackage[normalem]{ulem}
\usepackage[inkscapelatex=false]{svg}

\begin{document}

\title{Learning Context-Aware Service Representation for Service Recommendation in Workflow Composition
}

\author{\IEEEauthorblockN{Xihao Xie, Jia Zhang*}
\IEEEauthorblockA{\textit{*Department of Computer Science, Southern Methodist University, USA} \\\{xihaox, jiazhang\}@smu.edu}
}

\author{\IEEEauthorblockA{Xihao Xie\textsuperscript{1}, Jia Zhang\textsuperscript{1}, Rahul Ramachandran\textsuperscript{2},Tsengdar J. Lee\textsuperscript{3}, Seungwon Lee\textsuperscript{4}}
\IEEEauthorblockA{\textit{\textsuperscript{1}Department of Computer Science, Southern Methodist University, USA
} \\
\textit{\textsuperscript{2}NASA/MSFC, USA} \\
\textit{\textsuperscript{3}Science Mission Directorate, NASA Headquarters, USA}\\
\textit{\textsuperscript{4}NASA/JPL, USA} \\
\{xihaox,jiazhang\}@smu.edu;\{rahul.ramachandran,tsengdar.j.lee\}@nasa.gov;seungwon.lee@jpl.nasa.gov
}
}

\maketitle

\begin{abstract}
As increasingly more software services have been published onto the Internet, it remains a significant challenge to recommend suitable services to facilitate scientific workflow composition. This paper proposes a novel NLP-inspired approach to recommending services throughout a workflow development process, based on incrementally learning latent service representation from workflow provenance. A workflow composition process is formalized as a step-wise, context-aware service generation procedure, which is mapped to next-word prediction in a natural language sentence. Historical service dependencies are extracted from workflow provenance to build and enrich a knowledge graph. Each path in the knowledge graph reflects a scenario in a data analytics experiment, which is analogous to a sentence in a conversation. All paths are thus formalized as composable service sequences and are mined, using various patterns, from the established knowledge graph to construct a corpus. Service embeddings are then learned by applying deep learning model from the NLP field. Extensive experiments on the real-world dataset demonstrate the effectiveness and efficiency of the approach. \end{abstract}


\section{Introduction}
In recent years, increasingly more software programs have been deployed and published onto the Internet as reusable web services, or so-called APIs. Scientific researchers can thus leverage and compose existing services to build new data analytics experiments, so-called scientific workflow or workflow in short \cite{b1}.

From a practical perspective, exploiting existing services instead of reinventing the wheel will lead to higher efficiency. However, our studies over the life science field \cite{b2}, which is one of the very fields that adopt the concept of software service, revealed that the reusability rate of life science services in workflows remains rather low. Until 2018, less than 10\% of life science services published at biocatalog.org \cite{b3} were ever reused in scientific workflows, according to myExperiment.org \cite{b4}. This means that most of the published life science services have never been reused by anyone. 




\begin{figure}[htbp]
\centerline{\includegraphics[width=0.45\textwidth]{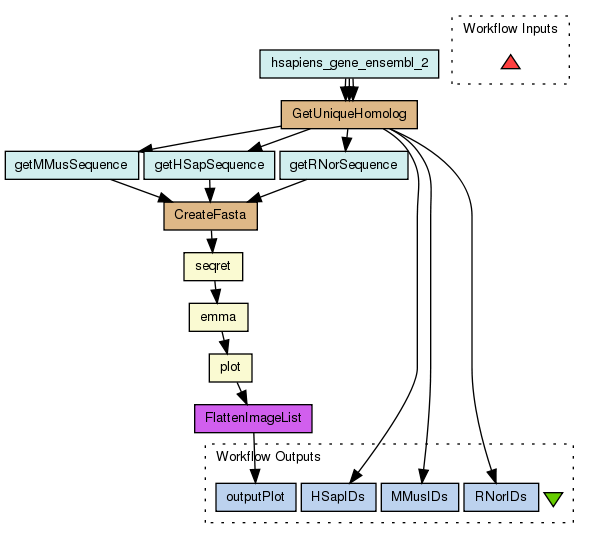}}
\caption{Motivating example workflow \#1794 in myExperiment.org}
\label{fig_motivating_example}
\end{figure}

What is wrong? Why scientists do not like to leverage others' work? Earlier studies believe one of the major obstacles making researchers unwilling to reuse existing services is the data shimming problem \cite{b5} \cite{b6}, meaning preparing and transforming data types to feed in the required inputs of a downstream service. For example, a NASA Climate Model Diagnostic Analyzer (CMDA) service, 2-D Variable Zonal Mean service that generates a graph of a 2-dimensional variable's zonal mean with time averaging, requires more than a dozen of input parameters \cite{b5}. Unless a user fully understands each of the parameters, both of its syntactic and semantic meaning and requirements, he/she may not feel comfortable to reuse the service. 

Our research project moves one step further, to argue that the adoption of a software service is not only dependent on its direct upstream service, but also the context of all selected services in the current workflow under construction. Take the workflow \#1794\footnote{https://www.myexperiment.org/workflows/1794.html} from myExperiment.org in Fig. \ref{fig_motivating_example} as an example. As highlighted in red oval, three services \textit{seqret}, \textit{emma}, and \textit{plot} are used sequentially to fetch a sequence set, do multiple alignments, and then plot the results. The adoption of the third service \textit{plot} aims to visualize the outcome from the two sequential services upstream in the workflow.

To tackle the shimming problem in the context of  service recommendation, our earlier studies constructed a knowledge graph \cite{b4}\cite{b7} from all historical data of service invocations, i.e., service provenance, extracted from workflow repositories such as myExperiment.org. Various random walk-powered algorithms were developed to traverse through the knowledge graph and suggest possible service candidates during workflow development \cite{b4}\cite{b7}.

In order to add into consideration of workflow-contextual information for more precise service recommendation in a recommend-as-you-go manner, this research presents a novel machine learning technique over service provenance knowledge graph. Inspired by the latest advancements in Natural Language Processing (NLP), we formalize the problem of service recommendation as a problem of next service prediction, where services and workflows are considered as ``tokens" and ``sentences" in NLP, respectively. In this way, our goal has turned into predicting and recommending the next suitable services that might be used for a user during the process of service composition.

The topic of next item prediction and recommendation has been well studied, and the literature has witnessed many successful applications in real-world fields such as e-commerce \cite{b9} \cite{b10}, keyboard prediction \cite{b11} and sequential click prediction \cite{b12}. In general, two distinct categories of approaches exist to recommend next items in a sequential context. Traditional approaches are typically based on Markov chain (MC). For example, Rendle et al. \cite{b13} modeled sequential data by learning a personalized transition graph over underlying MC to predict and recommend items that the user might want to purchase. In recent years, increasingly more deep learning-based approaches have been proposed for next item recommendation \cite{b9} \cite{b14}. Particularly, in NLP, the Word2Vec \cite{b15} model has achieved great success in learning the probability distribution of words to predict potential context based on sequential sentences. 

Inspired by Word2Vec, we leverage the skip-gram model to learn latent representation of services and their relationships for context-aware workflow recommendation. A corpus of ``sentences" (i.e., service chains extracted from historical workflow provenance) are generated from the knowledge graph. Our rationale is that, each path (i.e., a sequence of services) in a workflow reflects a scenario of data analytics experiment, which is analogous to a sentence in a conversation. Thus, all paths are extracted from the knowledge graph, analogous to all sentences are accumulated to train a computational model from NLP. In other words, a service sequence carries context when a service was invoked in the past.

To the best of our knowledge, we are the first attempt to seamlessly exploit the state-of-the-art machine learning techniques in both NLP and knowledge graph to facilitate service recommendation and workflow composition. Our extensive experiments over real-world dataset have demonstrated the effectiveness and efficiency of our approach. In summary, our contributions are three-fold:

\begin{itemize}
\item We formalize the service recommendation problem as a problem of context-aware next service prediction.
\item We develop a technique to generate a corpus of service-oriented-sentences (service sequences) by traversing over the knowledge graph established from workflow and service usage provenance.
\item We develop an approach to learning service representations offline based on sequential context information in accumulated knowledge graph and then recommending potential services real-time.
\end{itemize}

The remainder of this paper is organized as follows. Section \uppercase\expandafter{\romannumeral2} discusses the related work. In section  \uppercase\expandafter{\romannumeral3}, we present our context-aware service recommendation technique. In section  \uppercase\expandafter{\romannumeral4}, we present various ways of service chain corpus generation. Then, we discuss and analyze the experimental results in section \uppercase\expandafter{\romannumeral5}. Finally, section \uppercase\expandafter{\romannumeral6} draws conclusions.

\section{Related Work}
Our work is closely related to two categories of research in the literature: service recommendation and representation learning.

\subsection{Service Recommendation}
Service composition remains a fundamental research topic in services computing community. Paik et al.  \cite{b16} decomposed service composition activities into four phases: planning, discovery, selection and execution. Service recommendation represents a core technology in the third phase.

Zhang et al. \cite{b7} modeled services, workflows and their relationships from historical usage data into a social network, to proactively recommend services in a workflow composition process. In their later work \cite{b4}, they developed an algorithm to extract units of work (UoWs) from workflow provenance to recommend potentially chainable services. Chowdhury et al. \cite{b17} recommended composition patterns based on partial mashups. IBM's MatchUp \cite{b18} and MashupAdvisor \cite{b19} recommended reusable workflow components based on user context and conditional co-occurrence probability, respectively. By leveraging NLP techniques, Xia et al. \cite{b20} proposed a category-aware method to cluster and recommend services for automatic workflow composition. Shani et al. \cite{b8} formalized the problem of generative recommendation as a sequential optimization problem and applied Markov Decision Processes (MDPs) to solve it.

In Business Process Management (BPM) community, a number of research work have explored methods to recommend reusable components for workflow composition. VisComplete \cite{b21} recommended components in a workflow as a path extension by building graphs for workflows. Deng et al. \cite{b22} extracted relation patterns between activity nodes from existing workflow repository to recommend extending activities. Zhang et al. \cite{b23} mined the upstream dependency patterns for workflow recommendation. Similarly, Smirnov et al. \cite{b24} specified action patterns using association rule mining to suggest additional actions in process modeling.

\subsection{Representation Learning}
Representation learning, also known as feature learning, is a fundamental step for knowledge mining and downstream applications. Data representations matter for general data-driven algorithms and activities \cite{b25}. Many research communities have leveraged machine representation learning techniques on diverse data types to support real-world applications such as speech recognition \cite{b26}, signal processing \cite{b27}, language modeling \cite{b28} and semantic web \cite{b29}.

\begin{figure*}[htbp]
\centerline{\includegraphics[width=0.98\textwidth]{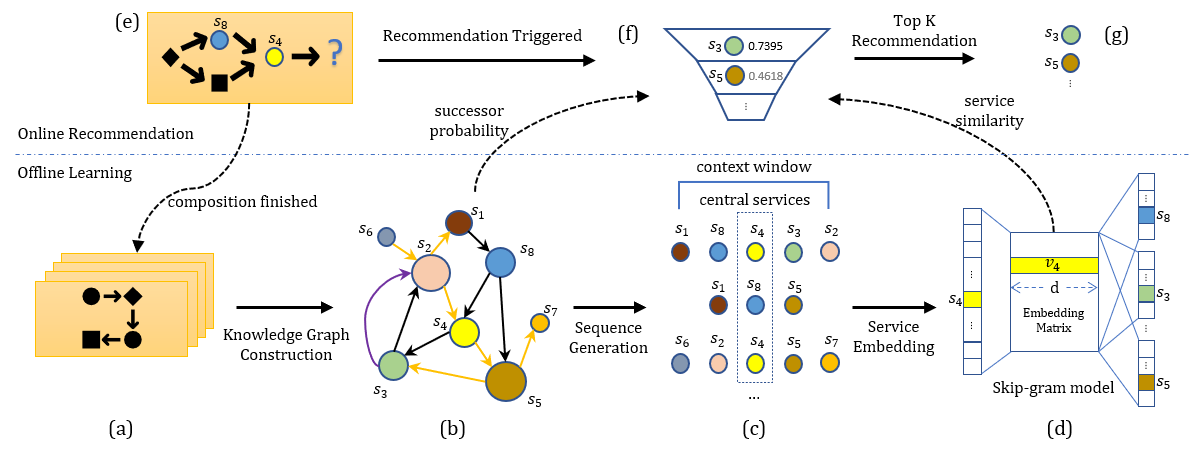}}
\caption{Blueprint of proposed approach. (a) Workflow repository. (b) Constructed knowledge graph. (c) Generated service token sequences according to different generation strategies. (d) Trained Skip-gram model offline. (e) Real-time workflow under construction. (f) Potential services with the product of successor probability and service similarity in descending order online. (g) Top-K recommended candidate services list. Operations from (a) to (d) are conducted in the offline training phase and (e)-(f) is the online recommendation phase.}
\label{fig_architecture}
\end{figure*}

In the services computing field, earlier work mainly focused on learning service representation based on natural language texts of service profiles. Semantic web is one of the technologies to build representations for service functions and properties. Li et al. \cite{b30} proposed an approach to recommending services by analyzing semantic compatibility between user requirements and service descriptions. Since Latent Dirichlet Allocation (LDA) was proposed in \cite{b31}, topic modeling has become a widely used method for learning service representation. Zhong et al. \cite{b32} applied the Author-Topic Models \cite{b33} to extract words as service description from mashup profiles. Zhang et al. \cite{b34} developed a tailored topic model to learn service representation for accurate visualization. Service2vec \cite{b41} constructed a service network \cite{b42} and applied Word2Vec modeling technique to extract contextual similarity between web services.

In contrast, our work differs from current literature of service representation and recommendation in three significant aspects. First, we believe that service usage context hides in paths in workflow provenance. Second, we formalize the problem of service recommendation as a problem of next service token prediction in sequential context and have explored different sequence generation strategies. Third, from another perspective of service representation, we learn service representation from the context and sequential dependencies of services in workflows instead of profile descriptions of services.

\section{Context-Aware Service Recommendation Approach}
In this section, we first depict the blueprint of our method. Then we formalize the research problem, followed by knowledge graph construction, offline service representation learning and online service recommendation. 

Fig. \ref{fig_architecture} presents the high-level blueprint of our methodology. Based on workflow provenance (a), we construct a knowledge graph (b) by extracting service dependencies from workflow structure. Second, we generate sequences of service tokens (c) from the knowledge graph. Third, we use language modeling techniques to learn latent representations of services (d). As shown in Fig. \ref{fig_architecture}, the above steps are conducted offline. The results of the offline learning phase will support the online recommendation at run-time. Given a workflow which is under composition online (e), we rank the potential services respect to their semantic similarities in the context (f), and recommend top-K of them (g) to the user at real time. Note that once a new workflow is completely composed, it will be saved into the workflow repository thus to trigger incremental offline learning.

\subsection{Problem Definition}
We consider the problem of service recommendation as a problem of predicting next service token, over a corpus of service sequences. The methods to generating service sequences from the knowledge graph, as shown in Fig. \ref{fig_architecture}, will be discussed in detail in the next section.

Let $\mathcal{W} = \left\{ w_{1}, w_{2}, ..., w_{\left| \mathcal{W} \right|} \right\} $ denote a set of workflows, $\mathcal{C} = \left\{ c_{1}, c_{2}, ..., c_{\left| \mathcal{C} \right|} \right\} $ be a set of available software components, i.e., services, $\mathcal{T} = \left\{ t_{1}, t_{2}, ..., t_{\left| \mathcal{T} \right|} \right\}$ be a set of service tokens. Note that each service token represents a subset of services, where an example may be a unit of work (several services always used together) \cite{b4}. In other words, the concept of service token is analogous to word phrases in NLP. Let list $S_{w'} = \left[s^{w'}_{1}, ..., s^{w'}_{k}, ..., s^{w'}_{|S_{w'}|} \right]$ denote a service sequence with length as $\left| S_{w'} \right|$ for workflow $w' \notin \mathcal{W}$ which has not been completed yet, where $s^{w'}_{k}$ is the service token $t_{k} \in \mathcal{T}$  generated at time step $k$ and $t_{k} \subseteq \mathcal{C}$ is a subset of services. 
Given a specific existing service sequence $S_{w'}$ of a workflow $w'$ under construction, our goal is to calculate the probability over all service tokens $t \in \mathcal{T}$ at next time step $\left| S_{w'} \right| + 1$:
\begin{equation}
p_{t} = p(t^{w'}_{\left|S_{w'}\right|+1}=t|S_{w'})
\label{eq_1}
\end{equation}
and recommend a tokens list $T \subseteq \mathcal{T}$ with top highest $n$ probability values, where $n$ is the size of the list $T$.

\subsection{Knowledge Graph Construction}
In this subsection, we introduce the details of constructing the web service knowledge graph (WSKG) from workflow repository, which will serve as the foundation for service representation learning and recommendation. Similar to our earlier method described in detail in \cite{b7}, WSKG is a directed graph (digraph) carrying historical service invocation dependencies extracted from workflow repository, and it is defined as:
\begin{equation}
WSKG = \left<\mathcal{C}, \mathcal{R}\right>
\end{equation}
where each software service $c \in \mathcal{C}$ is regarded as an entity, and $\mathcal{R} = \left\{r^{w}_{i,j}\right\}$ is a set of relationships between service entities. $r^{w}_{i, j} = \left<c_{i}, c_{j}, w\right>$ refers to a relationship that $c_{i}$ is an upstream service of  $c_{j}$ in workflow $w$. We can regard the relationship $r^{w}_{i,j}$ as an edge starting from $c_{i}$ and ending at $c_{j}$ with label $w$. Note that there might be multiple edges between two service nodes in the knowledge graph with different labels, meaning that such a service invocation dependency happens in multiple workflows.

As shown in Fig. \ref{fig_architecture}, the constructed WSKG serve for both the offline learning phase and the online recommendation phase. For the offline phase, the  constructed WSKG allows us to generate all service token sequences, by traversing the WSKG following the comprising paths. In section IV, we will further discuss various strategies of generating service token sequences in WSKG. All service token sequences together will form a corpus to learn latent service token representations. 

In the online recommendation phase, the structural relationships carried in WSKG can be leveraged to enhance service ranking. For example, such information will help empirically model the probability of service token $t$ appearing as a successor of service $c$ in a partial workflow under development.

\subsection{Offline Service Representation Learning}
The offline representation learning phase aims to learn latent service representations over the service sequence corpus. Two models, skip-gram and CBOW in Word2Vec, are widely applied neural network models in NLP to learn word representations given a corpus of sentences \cite{b15}. In our study, we decide to employ the skip-gram model mainly because it works well with small amount of training data and represents well even for rare tokens \cite{b43}. In our scenarios, existing scientific workflow datasets are not so large as that in the field of NLP. Furthermore, the significant amount of long-tail services (e.g., newly published services) may be able to receive reasonable exposure using the skip-gram model, although they rarely appear in the service sequence corpus.

Applying the skip-gram model \cite{b15}, we turn our formalized problem \eqref{eq_1} into an optimization problem of maximizing the probability of any service token appearing in the on-going service sequential context:
\begin{equation}
\mathop{max}\limits_{\Omega}\sum\limits_{t_{i}\in\mathcal{T}}\log{p\left(S_{t_{i}} | \Omega\left(t_{i}\right) \right)}
\label{eq_2}
\end{equation}
where $S_{t_{i}} = \left[t_{i - w}, ..., t_{i - 1}, t_{i + 1}, ..., t_{i + w}\right]$ is the sequential context of token $t_{i} \in \mathcal{T}$, $w$ is the window size and $\Omega\left(t_{i}\right) \in \mathbb{R}^{d}$ is the learnt service representations in form of vectors.

Based on an independence assumption, the probability of \eqref{eq_2} can be approximated as:
\begin{equation}
{p\left(S_{t_{i}} | \Omega\left(t_{i}\right) \right)} = \prod_{j = i - w,j \neq i}^{i + w}p\left(t_{j} | \Omega\left(t_{i}\right)\right)
\label{eq_3}
\end{equation}

Since computing $p\left(t_{j} | \Omega\left(t_{i}\right)\right)$ is time consuming and hardly feasible, we decide to employ Hierarchical Softmax \cite{b35} to speed up the training process. Specifically, we build a Huffman tree whose leaves are service tokens. For any leaf node $n_{k}$, one path $\left[n_{0}, n_{1}, ..., n_{h_{k}}\right]$ exists from the root $n_{0}$ to it, where $h_{k}$ is the length of the path and limits to $\lceil \log\left|\mathcal{T}\right| \rceil$. In this way, we can approximate the probability in \eqref{eq_3} as:
\begin{equation}
p\left(t_{j} | \Omega\left(t_{i}\right)\right) = \prod_{k = 1}^{h_{k}}p\left(n_{k} | \Omega\left(t_{i}\right) \right)
\label{eq_4}
\end{equation}

As for $p\left(n_{k} | \Omega\left(t_{i}\right) \right)$, we view it as a binary classification problem and calculate it as in \eqref{eq_5}:
\begin{equation}
p\left(n_{k} | \Omega\left(t_{i}\right) \right) = \frac{1}{1 + \exp(-\Omega^\mathrm{T}\left(n_{k}\right) \cdot \Omega\left(t_{i}\right))}
\label{eq_5}
\end{equation}

As a result, the time complexity of computing $p\left(t_{j} | \Omega\left(t_{i}\right)\right)$ can be reduced from $O\left(\left|\mathcal{T}\right|\right)$ to $O\left(\log\left|\mathcal{T}\right|\right)$. Algorithm 1 illustrates the step-wise procedure of offline representation learning. Note that during the training process, we update vectorized representations with stochastic gradient descent, as shown in lines 6 and 7.

\subsection{Online Service Recommendation}
As shown in Fig. \ref{fig_architecture}, at each step of the process of composing a new workflow, service token $t_{l}$ in current user session can be used as the input to trigger recommendation for the next service token. After that, the online recommendation engine ranks all potential service tokens according to their similarities to $t_{l}$ and recommend the top-$K$ of them as candidates to the user. The user will then select a service token from the recommended list to move the composition process forward. Note that a service token in our work might be a single service or a bundle of services due to different sequence generation strategies, which will be discussed in the next section. That means each entry of the recommended list might contain one or more services. If an entry is comprised of multiple services, the user can further select one or more services from the entry to continue the composition. A good analogy of a token containing multiple services is a \emph{phrase} in natural language. In this case, recommending a group of services is like predicting the most probable next phrase in language writing \cite{b11}.

Following the skip-gram model, given a specific sequence of service tokens $\left[t_{l - u}, ..., t_{l - 1}, t_{l}\right]$, the offline learnt service representations enable us to identify the top-$K$ most \emph{relevant} service tokens to be the most potential service tokens following $t_{l}$. According to the skip-gram model, though, the predicted contextual service tokens may not only be upstream of $t_{l}$ in some workflows, but also be downstream service tokens as well. However, in the scenario of workflow composition, what we need to recommend is the service tokens that are potentially appear after $t_{l}$. Therefore, we define a scoring function which is used to rank next potential service tokens in descending order as follows:
\begin{equation}
score(t_{l}, t) = p_{suc}(t_{l}, t) \times sim(t_{l}, t)
\end{equation}
\begin{equation}
p_{suc}(t_{l}, t) = \frac{\exp(N_{suc}(t_{l}, t))}{\exp(N_{pre}(t_{l}, t)) + \exp(N_{suc}(t_{l}, t))}
\end{equation}
where $p_{suc}(t_{l}, t)$ empirically models the probability that $t$ appears after $t_{l}$, $N_{suc}(t_{l}, t)$, $N_{pre}(t_{l}, t)$ are the numbers of occurrences that any service $c \in t$ appeared in repository as a successor or a precursor service of $t_{l}$, respectively. $sim(t_{l}, t)$ is the similarity between $t_{l}$ and $t$ that can be calculated from the service representations learnt offline.


\section{Service Sequences Generation}
The input of the used skip-gram representation learning model is a corpus of service sequences, each of which is comprised of service tokens. As mentioned above, how to generate sequences of service tokens from WSKG is the key step in our method. Different generation strategies may result in different recommendation. In this section, we explore three different ways, each of which might be adapt to a specific scenario, to generate sequential service tokens from the constructed knowledge graph WSKG that is introduced in the previous section. 

\begin{figure*}[htbp]
\centering
\includegraphics[width=0.8\textwidth]{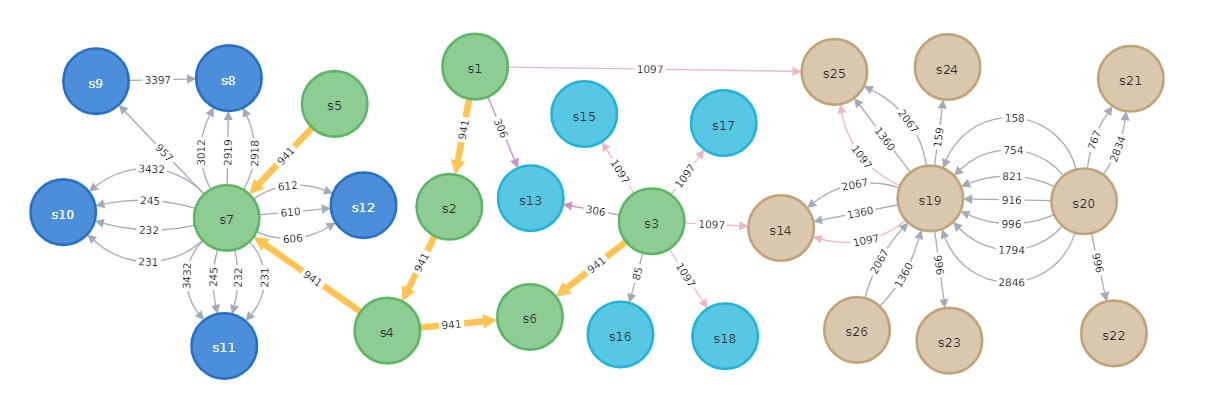}
\caption{Portion of WSKG motivating three service sequence generation methods. The nodes in green are services in workflow \#941 and the dependencies between them are colored in orange with label ``941." Nodes in other colors are services invoked by other workflows, which appear to be downstream nodes of the services in workflow \#941 in WSKG.}
\label{fig_tangible_workflow}
\end{figure*}

The three generation methods are: DFS-based generation, BFS-based generation, and PW-based generation. Fig. \ref{fig_tangible_workflow} is a portion extracted from the constructed WSKG that motivates and will be used to explain the three service sequence generation methods. The nodes in green represent the services in workflow \#941\footnote{https://www.myexperiment.org/workflows/941.html} published at myExperiment.org. Nodes in other colors illustrate some downstream services from other workflows recorded in WSKG. Edges are labeled with corresponding id numbers of workflows.


\begin{algorithm}[H]
\caption{Offline Learning Service Representation}
\begin{algorithmic}[1]
\renewcommand{\algorithmicrequire}{\textbf{Input:}}
\renewcommand{\algorithmicensure}{\textbf{Output:}}
\REQUIRE workflows $\mathcal{W}$, windows size $w$, dimension size $d$ and sequence generation type $\mbox{T}$
\ENSURE vectorised service representations $\Omega \in \mathbb{R}^{\left|\mathcal{C}\right| \times d}$
\STATE Initialize $\Omega \in \mathbb{R}^{\left| \mathcal{C} \right| \times d}$
\STATE $\mathcal{D} \gets GenerateSequences(\mathcal{W}, \mbox{T})$
\\$//$ \small{generate service token sequences according to specific generation type \mbox{T}, see in next section.}
\FOR{each sequence $S \in \mathcal{D}$ }
  \FOR{each token $t_{i} \in S$ }
    \FOR{each $t_{j} \in S\left[i - w: i + w\right]$}
      \STATE $J\left(\Omega\right) = -\log p\left(t_{j}|\Omega{\left(t_{i}\right)}\right)$
      \STATE $\Omega = \Omega - \eta \times \frac{\partial J}{\partial \Omega}$ // $\eta$ is the learning rate.
    \ENDFOR
  \ENDFOR
\ENDFOR
\label{alg_offline_learning}
\end{algorithmic}
\end{algorithm}

\subsection{Depth First Search (DFS) based Generation}
In this method, we consider a workflow composition process as a path extension process from start to end. Particularly, we consider individual workflows. In order to model the sequential behaviour of composing a specific workflow $w \in \mathcal{W}$, depth first search method is applied in WSKG, along the path with the label $w$ to generate sequences of service tokens. Specifically, for any service entity $c_{i} \in \mathcal{C}$, let $S^{w}_{i, m} = t^{i}_{1} \xrightarrow[]{w} t^{i}_{2} \xrightarrow[]{w} ... \xrightarrow[]{w} t^{i}_{\left| S^{i}_{w} \right|}$ denote a generated sequence starting from $c_{i}$ 
and going along the path with label $w$, representing a workflow, to next unvisited neighbor services until meeting terminal services which have no successor services along label $w$. Here $t^{i}_{1} = c_{i}$, $t^{i}_{j} \in \mathcal{T}$ and $\forall{t^{i}_{j}}$ has $\left|t^{i}_{j}\right| = 1$, which means every token is a single service. $m \in \left[1, M\right]$ and $M$ is the total number of sequences starting from $c_{i}$. 

Note that for a workflow with label $w$, the sequences are generated from not only its starting services but also all intermediate services. In this way, the generated sequences can cover as many as possible sequential dependencies among services in the workflow $w$.


Take the workflow \#941 from myExperiment for a simple example. For illustration purpose, Fig. \ref{fig_tangible_workflow} uses $s_{1}$, $s_{2}$, ..., $s_{7}$ to stand for the services nodes of \emph{String\_Constant}, \emph{getPeak\_input}, \emph{String\_Constant1}, \emph{getPeak}, \emph{String\_Constant0}, \emph{XPath\_From\_Text0}, and \emph{XPath\_From\_Text}, respectively. In the corresponding workflow, the nodes $s_{1}$, $s_{3}$, $s_{5}$ are starting service entities and nodes $s_{6}$, $s_{7}$ are terminal service entities that can be traversed along the label ``941." Applying the DFS based generation strategy, we can firstly generate four services sequences, two of which starting from $s_{1}$: $S^{941}_{1, 1} = s_{1}\xrightarrow[]{941}s_{2}\xrightarrow[]{941}s_{4}\xrightarrow[]{941}s_{6}$, $S^{941}_{1, 2} = s_{1}\xrightarrow[]{941}s_{2}\xrightarrow[]{941}s_{4}\xrightarrow[]{941}s_{7}$, $S^{941}_{3, 1} = s_{3}\xrightarrow[]{941}s_{6}$ and $S^{941}_{5, 1} = s_{5}\xrightarrow[]{941}s_{7}$. Afterwards, four more sequences can be generated starting from intermediate services: $S^{941}_{2, 1} = s_{2} \xrightarrow[]{941} s_{4} \xrightarrow[]{941} s_{6}$, $S^{941}_{2, 2} = s_{2} \xrightarrow[]{941} s_{4} \xrightarrow[]{941} s_{7}$, $S^{941}_{4, 1} = s_{4} \xrightarrow[]{941} s_{6}$, $S^{941}_{4, 2} = s_{4} \xrightarrow[]{941} s_{7}$.

\subsection{Breadth First Search (BFS) based Generation}
The main idea of the BFS based service tokens generation lies in multiple items recommendation. In order to achieve higher business goal, a real-world recommender system typically not only recommend single items but also a bundle of items. Actually, recommending next bundle of services can be regarded as recommending next basket of items that a user might want to buy in a single visit in e-commerce scenarios \cite{b13}, \cite{b9}. Our earlier research also studied how to recommend unit of work (UoW) with a collection of services usually used together, based on network analysis \cite{b4}.

For the similar reason, we consider service bundles as services tokens when generating sequential services tokens from WSKG. Applying BFS based generation strategy, for any service $c_{i} \in \mathcal{C}$ in a workflow $w$, the sequence of service tokens starting from $c_{i}$ can be generated as: $S^{w}_{i} = \left[t^{i}_{1}, t^{i}_{2}, ..., t^{i}_{\left|S^{w}_{i}\right|}\right]$, where $t^{i}_{k} \subseteq \mathcal{C}$ is a set of multiple services that are at the $k$\textsuperscript{th} level of the breadth first searching path. Specially, $t^{i}_{1}$ is a set containing only $c_{i}$, and it can be either a starting service or an intermediate service.

As for the sub-WSKG in Fig. \ref{fig_tangible_workflow}, for example, we can generate three service sequences starting from $s_{1}$, $s_{3}$ and $s_{5}$: $S^{941}_{1} = \left[s_{1}, s_{2}, s_{4}, s_{6}\&s_{7}\right]$, $S^{941}_{3} = \left[s_{3}, s_{6}\right]$ and $S^{941}_{5} = \left[s_{5}, s_{7}\right]$. We can see that $s_{6}$ and $s_{7}$ are combined together as the fourth token in the sequence of $S^{941}_{1}$. It means that any permutation of them can be composed following the service in the third step of the workflow. Sequences starting from other intermediate services can be generated similarly: $S^{941}_{2} = \left[s_{2}, s_{4}, s_{6}\&s_{7}\right]$, $S^{941}_{4} = \left[s_{4}, s_{6}\&s_{7}\right]$, $S^{941}_{6} = \left[s_{6}\right]$, $S^{941}_{7} = \left[s_{7}\right]$. For label $``1097"$ in Fig. \ref{fig_tangible_workflow}, we can generate another sequence $S^{1097}_{3} = \left[s_{3}, s_{14}\&s_{15}\&s_{17}\&s_{18}\right]$. For label $``306,"$ we can generate two sequences starting from $s_{1}$ and $s_{3}$, respectively: $S^{306}_{1} = \left[s_{1}, s_{13}\right]$, $S^{306}_{3} = \left[s_{3}, s_{13}\right]$.


\subsection{Probabilistic Walk (PW) based Generation}
Inspired by \textsc{DeepWalk} \cite{b37}, which uses random walk to generate sequences of nodes to learn latent representations for vertices in a graph, we leverage a variant of random walk, i.e., probabilistic walk, to generate sequences of services based on the constructed knowledge graph WSKG.

Recall that the DFS and BFS based generation algorithms discussed earlier both generate a service sequence starting from service token $t_{j}$ along the path with a specific label $w$. In contrast, the probabilistic walk based sequence generation algorithm considers the generation of next service token rooted at $t_{j}$ as a stochastic process, with random variables $T^{1}_{j}, T^{2}_{j}, ..., T^{l}_{j}$ where $T^{l+1}_{j}$ is a service generated with probabilities from the neighbors of service token $t_{l}$. Note that, such a service token $T^{l+1}_{j}$ existing in the sequence of $\left[ T^{1}_{j}, T^{2}_{j}, ..., T^{l}_{j} \right]$ is not allowed, meaning that the generated sequence is acyclic. On the one hand, restarting from a vertex did not show any improvement to our experimental results. On the other hand, it is generally meaningless to invoke a previously invoked service in the same workflow. Specifically, for a service $u\in{\mathcal{C}}$, let $N_{u} \subseteq{\mathcal{C}}$ denote the set of directed neighbor services in the whole WSKG, we model the probability $p_{u,v}$ that service token $v$ can be generated after $u$:
\begin{equation}
p_{u,v} = p(v|u)=\sigma(\frac{o_{u,v}}{o_{u}})=\frac{\exp(\frac{o_{u,v}}{o_{u}})}{\sum_{n\in{N_{u}}}\exp(\frac{o_{u,n}}{o_{u}})}
\label{eq_puv}
\end{equation}
where $v\in{N(u)}$, $\sigma$ is the commonly used softmax function, $o_{u,v}$ is the number of occurrence of relationships between $u$ and $v$ in the whole WSKG and $o_{u} = \sum_{n\in{N(u)}}o_{u,n}$.


For example, in Fig. \ref{fig_tangible_workflow}, the dependency relationship between $s_{7}$ and $s_{10}$ occurs four times in four different workflows whose id numbers are ``3432," ``245," ``232" and ``231," respectively. However, for $s_{9}$, it appears after $s_{7}$ only once in workflow ``957". Therefore, according to Eq. \ref{eq_puv}, the probability that $s_{10}$ can be generated as a downstream service token of $s_{7}$ is larger than that of $s_{9}$.

Two factors may impact the effectiveness of the PW based generation strategy. One factor is the length of a visiting path, i.e., $l$, which determines when to stop while walking along a path. The other factor is the number of walks starting at each vertex, i.e., $\theta$. In practice, without specifying $\theta$, some linkages may not be traversed. As a result, the generated sequences may not cover all tangible dependency relationships, tampering the ability of trained representations to predict next suitable services. As \cite{b37} suggested, although it is not strictly required to do this, but it is a common practice to tune the two factors to speed up the convergence of stochastic gradient descent in algorithm 1. In the next section of experiments, We will discuss in detail the effects of $l$ and $\theta$.



\subsection{Discussions}
For all of the three aforementioned strategies, no matter which one we apply, there might appear duplicate service sequences. The reason is that some services and their dependency relationships in a workflow $w_{i}$ might exist in another workflow $w_{j}$, in a form of service chain. For example, in Fig. \ref{fig_tangible_workflow}, applying DFS-based strategy, sequence $s_{26} \xrightarrow[]{} s_{19} \xrightarrow[]{} s_{25}$ can be generated twice through two workflows, i.e., with labels ``1360" and ``2067." Similarly, with labels ``1097," ``1360" and ``2067," sequence $\left[s_{19}, s_{14}\&s_{25}\right]$ can be generated three times by applying BFS-based strategy.


As for the PW-based strategy, three factors may result in duplicates: the length of a path, the frequencies of the dependency relationships in the path and the predefined parameter $\theta$. A shorter path with more frequent dependency relationships tends to be generated as a sequence with higher probability in case of a higher $\theta$. For example, in Fig. \ref{fig_tangible_workflow}, let $s_{20}$, $s_{19}$ and $s_{14}$ stand for services \emph{seqret}, \emph{emma} and \emph{emma\_NJ} respectively. The frequency of the dependency relationship between $s_{20}$ and $s_{19}$ is much higher than the frequencies of relationships between $s_{20}$ and other services. So does that of the relationship between $s_{19}$ and $s_{14}$. Therefore, the sequence $s_{20} \xrightarrow[]{} s_{19} \xrightarrow[]{} s_{14}$, whose length is three, is generated multiple times.

Similar to insurmountable duplicates in a natural language corpus \cite{b36}, duplicates in the corpus of service token sequences generated from the knowledge graph might impact the effectiveness of our approach. It is a \emph{feature-but-not-bug} problem. We will evaluate the effect of duplicates with experiments in the next section.

\section{Experiments}
In this section, we first provide an overview of the dataset, then explain the evaluation metrics used and present the experimental results and analyses.

\subsection{Dataset}
Since 2007, myExperiment.org has become the largest service-oriented scientific workflow repository in the world. It has been used as a testbed by many researchers \cite{b4}, \cite{b5}, \cite{b7} in the services computing community. Since we construct the knowledge graph WSKG across workflow boundaries, we target on the Taverna-generated workflows, i.e., workflows generated following Taverna syntax.

In the target testbed, every workflow is maintained as an XML file and every service in a workflow is defined as a \textit{processor}. A dependency relationship between two services is defined as a \textit{link} or \textit{datalink} whose two ends are \textit{source} and \textit{sink}, respectively. We examined all Taverna workflows published on myExperiment.org up to October 2021, with a total of 2,910 workflows and 9,120 services. Table \ref{tab1} lists the summarized information over all workflows in the dataset.

\begin{table}[htbp]
\caption{Statistical Information about myExperiment Dataset}
\label{tab1}
\begin{center}
\begin{tabular}{|c|c|c|}
\hline
\multicolumn{2}{|c|}{Total \# of workflows} & 2,910 \\\hline
\multicolumn{2}{|c|}{Total \# of services}  & 8,837 \\\hline
& BFS & 10,788 \\
\cline{2-3}
Total \# of sequences & DFS & 41,163 \\
\cline{2-3}
& PW & 50,314 \\
\hline
& BFS & 5.0 \\
\cline{2-3}
Average length of sequences & DFS & 9.0 \\
\cline{2-3}
& PW & 7.0 \\
\hline
\end{tabular}
\end{center}
\end{table}

\subsection{Experimental Setup}
We randomly used 80\% of the workflows to generate service sequences as training data to learn service representations. The remaining 20\% of workflows were utilized as the test data. In the offline learning phase, we set the number of window size as 3, and the dimension size $d$ as 50. The generated service sequences with length smaller than 2 were removed from the training set. As for the online recommendation phase, similar to the \emph{leave-one-out} task that has been widely used in sequential recommendation \cite{b38}, \cite{b39}, \cite{b40}, we adopted the \emph{leave-last-service-out} cross-validation to evaluate our approach. 

For each workflow in the test set, we removed the terminate services of the workflow and used the services just before the terminate services as the inputs of the trained representation model to rank top $K$ service tokens as candidates. Hence, the experiments were turned to evaluate whether the recommended service tokens hit the ground truth. Additionally, in the test test, we removed the linkage which never occurred in the sequences generated from the training data, considering that the skip-gram model is not able to predict tokens that are not in the vocabulary.

\begin{figure*}[htbp]
\centerline{
\includegraphics[width=0.24\textwidth]{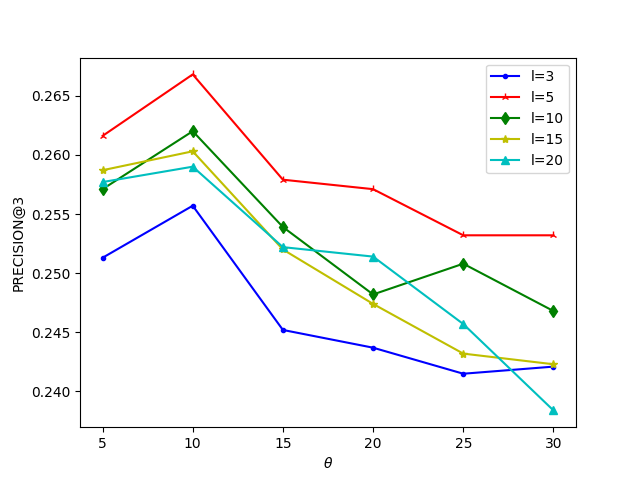}
\includegraphics[width=0.24\textwidth]{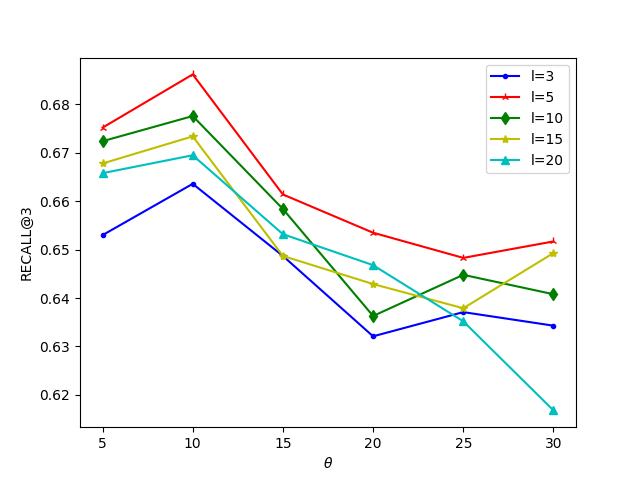}
\includegraphics[width=0.24\textwidth]{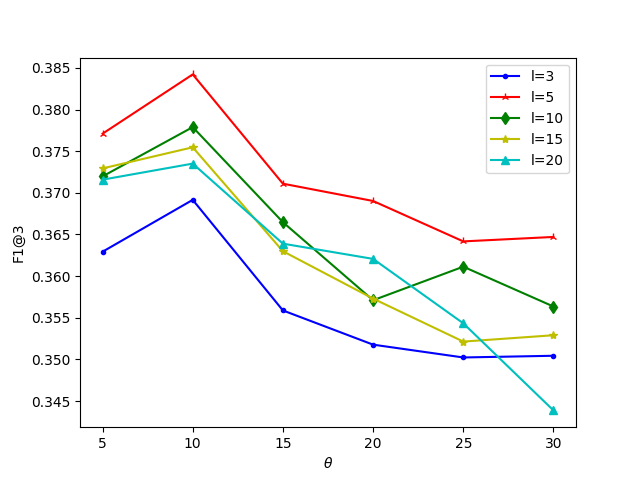}
\includegraphics[width=0.24\textwidth]{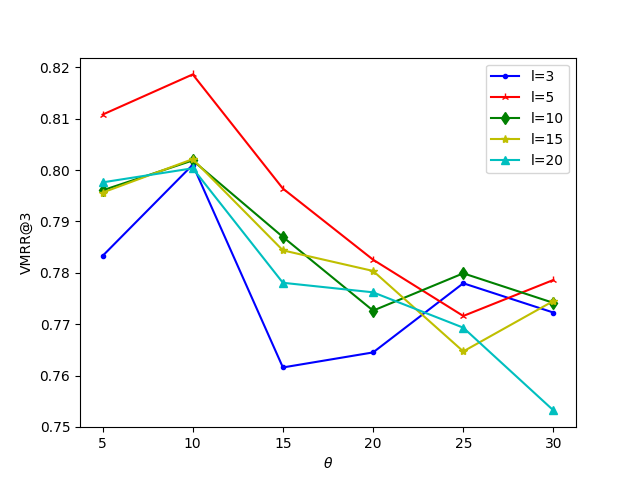}
}
\caption{The effect of $l$ and $\theta$ with top-3 recommendations}
\label{fig_parameter_sensitivity_3}
\end{figure*}

\begin{figure*}[htbp]
\centerline{
\includegraphics[width=0.24\textwidth]{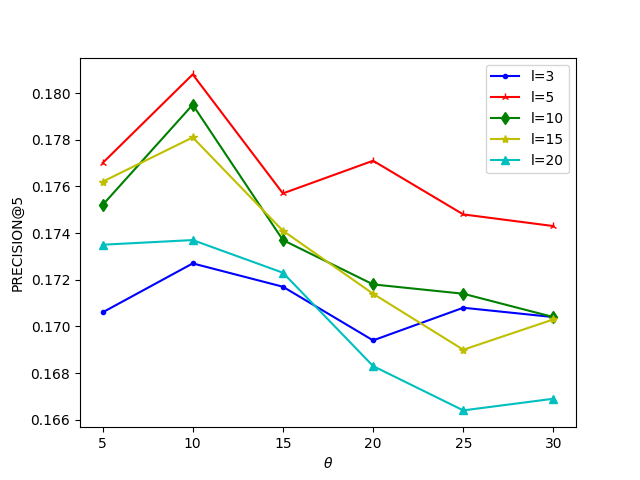}
\includegraphics[width=0.24\textwidth]{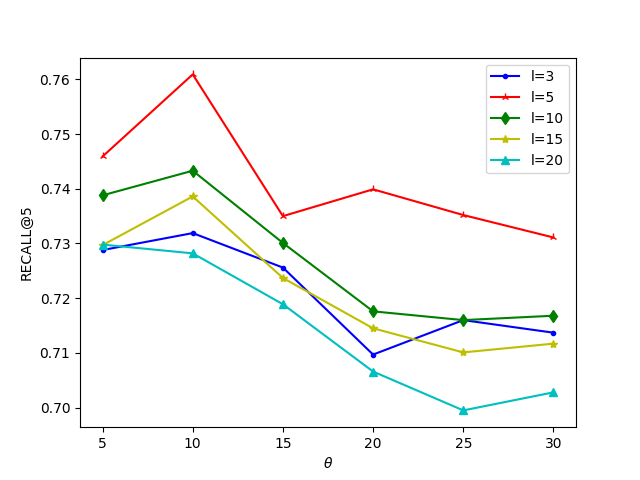}
\includegraphics[width=0.24\textwidth]{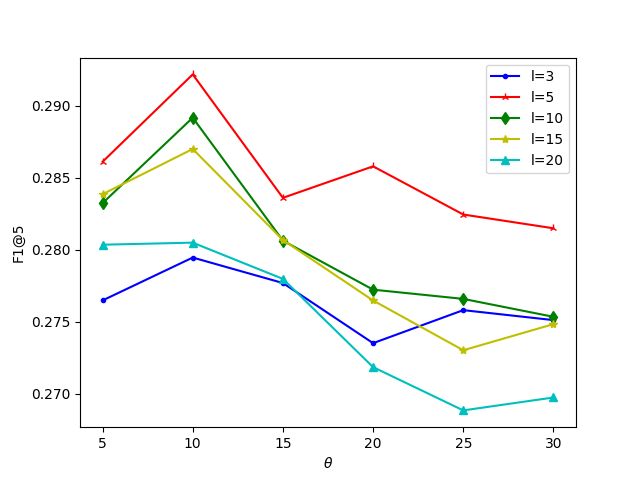}
\includegraphics[width=0.24\textwidth]{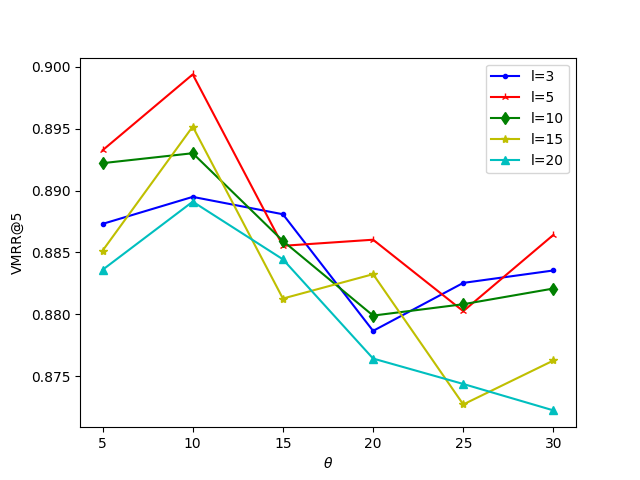}
}
\caption{The effect of $l$ and $\theta$ with top-5 recommendations}
\label{fig_parameter_sensitivity_5}
\end{figure*}

\begin{figure*}[htbp]
\centerline{
\includegraphics[width=0.24\textwidth]{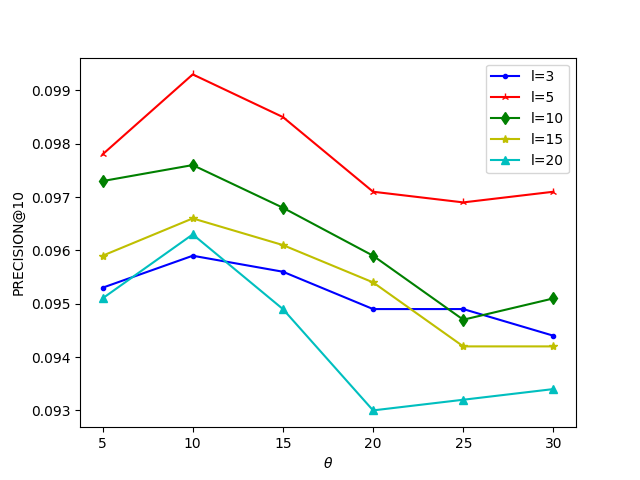}
\includegraphics[width=0.24\textwidth]{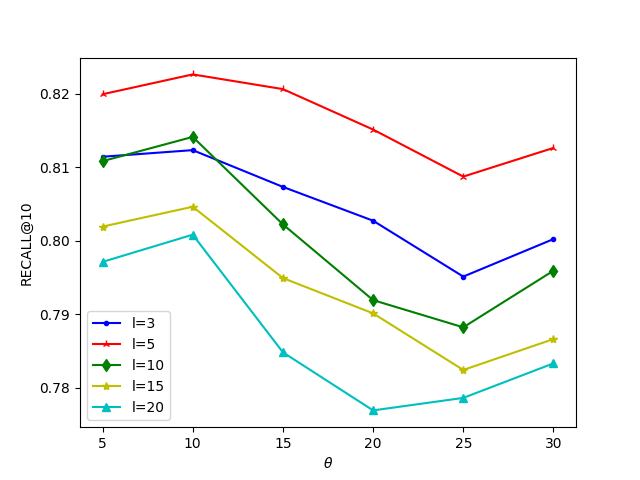}
\includegraphics[width=0.24\textwidth]{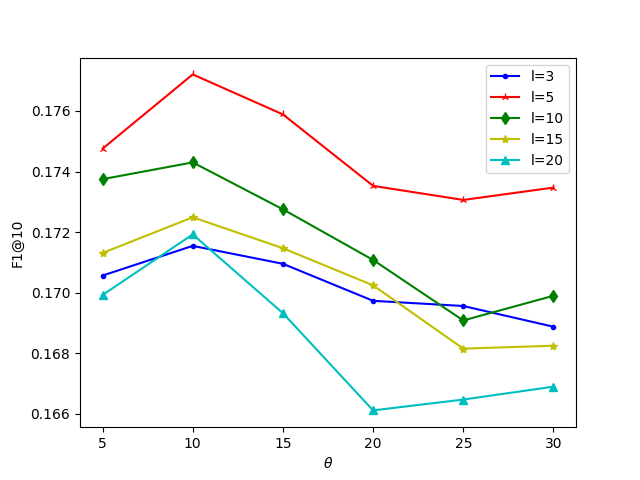}
\includegraphics[width=0.24\textwidth]{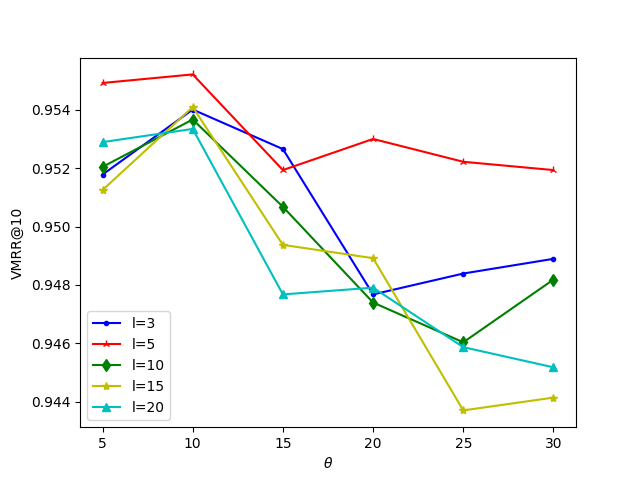}
}
\caption{The effect of $l$ and $\theta$ with top-10 recommendations}
\label{fig_parameter_sensitivity_10}
\end{figure*}

\subsection{Evaluation Metrics}
All of the evaluation metrics of our approach aim to evaluate whether the recommended top $K$ entries hit the ground truth or not. For DFS-based and PW-based generation strategy, each entry is a single service token. An entry $e_{i} \in R$ hits the ground truth $G$ if $e_{i} \in G$, where $R$ is the recommended result, and $\left|R\right| = K$. For BFS-based generation strategy, an entry $e_{i} \subseteq \mathcal{C}$ in the recommended list might be a service token comprised of multiple services. In this case, the entry $e_{i}$ hits the ground truth $G$ if $\exists c_{j} \in e_{i}$ such that $c_{j} \in G$.

To evaluate our recommendation approach, we employed \textit{PRE@K}, \textit{REC@K} and \textit{F1@K}, which are short for \textit{precision}, \textit{recall} and \textit{F1 Score}, respectively, as the evaluation metrics. For every workflow in the test data set, they can be calculated as follows:
\begin{equation}
precision = \frac{\left|R \cap G \right|}{\left|R\right|}
\end{equation}
\begin{equation}
recall = \frac{\left|R \cap G \right|}{\left|G\right|}
\end{equation}
\begin{equation}
F1 = 2 \cdot \frac{precision \cdot recall}{precision + recall}
\end{equation}
where $\left|R \cap G\right|$ means the number of hit entries in the top $K$ recommended result. Additionally, consider that the index of the hit result (i.e., \textit{idx}) in the recommended list is a significant factor to users' willingness to reuse recommended services. We thus introduced a variant of MRR, aka \textit{VMRR}, as an evaluation metric as well. It helps us to capture the overall performance of the rank of the hit entry in the recommended result. The calculation of \textit{VMRR} is:
\begin{equation}
VMRR = 
\begin{cases}
0 & R \cap G = \varnothing \\
\frac{\left|R\right| - idx}{\left|R\right|} = 1 -  \frac{idx}{K} &  else
\end{cases}
\end{equation}
For these four metrics, we reported their mean values over all workflows in the test dataset. The higher the value, the better the performance.


\begin{table*}[htbp]
\caption{Overall Recommendation Results}
\label{tab_overall_result}
\begin{center}
\begin{tabular}{c|c c c c|c c c c|c c c c}
\hline
Metrics & PRE@3 & REC@3 & F1@3 & VMRR@3 & PRE@5 & REC@5 & F1@5 & VMRR@5 & PRE@10 & REC@10 & F1@10 & VMRR@10 \\\hline
BFS & \underline{0.3059} & 0.6258 & \underline{0.4109} & 0.7651 & \underline{0.2191} & 0.6897 & \underline{0.3323} & 0.8730 & \underline{0.1031} & 0.7397 & \underline{0.1810} & 0.9412 \\
DFS & 0.2384 & 0.6142 & 0.3435 & 0.7516 & 0.1591 & 0.6721 & 0.2573 & 0.8612 & 0.0893 & 0.7455 & 0.1595 & 0.9362 \\
PW & 0.2564 & 0.6457 & 0.3609 & 0.7735 & 0.1668 & 0.6967 & 0.2692 & 0.8681 & 0.0912 & 0.7617 & 0.1629 & 0.9392  \\
BFS-DR & \textbf{0.3144} & \underline{0.6465} & \textbf{0.4231} & \underline{0.7789} & \textbf{0.2249} & \underline{0.6979} & \textbf{0.3402} & \underline{0.8779} & \textbf{0.1099} & 0.7579 & \textbf{0.1920} & \underline{0.9445} \\
DFS-DR & 0.2449 & 0.6243 & 0.3518 & 0.7607 & 0.1630 & 0.6835 & 0.2632 & 0.8668 & 0.0922 & \underline{0.7624} & 0.1645 & 0.9404 \\
PW-DR & 0.2668 & \textbf{0.6862} & 0.3842 & \textbf{0.8186} & 0.1808 & \textbf{0.7609} & 0.2922 & \textbf{0.8994} & 0.0993 & \textbf{0.8226} & 0.1772 & \textbf{0.9552} \\
\hline
\end{tabular}
\end{center}
\end{table*}

\subsection{Parameter Sensitivity}
We designed experiments to evaluate how changes to $l$ and $\theta$ will effect the performance of the probabilistic walk based generation strategy. In such experiments, duplicate generated sequences were removed. Figs. \ref{fig_parameter_sensitivity_3}, \ref{fig_parameter_sensitivity_5} and \ref{fig_parameter_sensitivity_10} show the effects of $l$ and $\theta$ to the PW-based generation strategy. In terms of precision, recall and F1, $l = 5$ is better than $l = 3$, $l = 10$, $l = 15$ and $l = 20$. Note that the median and mean lengths of workflows are 8 and 9.49, respectively, both of which are closed to 9. Approximately, for a workflow with length of 9, the average length of generated sequences with BFS-based generation strategy is around 5. A lower $l$ results in lower coverage over all linkages between services, which reduces the capability of discovering suitable services following a specific service. A higher $l$ makes it more likely to generate service sequences that are not existed in workflows.

The parameter $\theta$ in this work is similar to the parameter $\gamma$ in \cite{b37}. For $\theta$, 10 is almost the best option. The reason might be that, for all nodes in the constructed WSKG, their mean out-degree and mean number of downstream nodes are 2.38 and 1.69, respectively, both of which are close to 2.0. Approximately, for a binary tree with height of 5, to reach a specific leaf from the root with probabilities at forks, it needs about 10 attempts. 
On the one hand, a lower $\theta$ might result in higher probability to miss tangible dependencies that are useful for next service recommendation. On the other hand, a higher $\theta$ tends to generate duplicate sequences, which might be misleading for representation learning.

In summary, our experiments demonstrated that $l$ and $\theta$ matter for probabilistic walk based sequence generation strategy. Our suggestion is that when applying the PW-based strategy, it will be good to get a glimpse at the length of workflows and the out-degree of service nodes in the constructed graph.

\subsection{Performance Comparison}
We designed experiments to answer two research questions. First, which sequence generation strategy acts the best? Second, should duplicates from the generated sequences be removed? Table \ref{tab_overall_result} presents the overall recommendation performance metrics among six generation strategies. BFS-DR, DFS-DR and PW-DR are results of BFS, DFS and PW with duplicate sequences removed, respectively. For each metric, the best result is highlighted in bold and the second best is underlined. According to the lessons we learnt as discussed earlier, we set $l$ and $\theta$ to 5 and 10, respectively, for PW and PW-DR.

\emph{RQ1: Which sequence generation strategy is the best?} 
According to the experimental results, in terms of precision and F1, BFS-based strategies outperform others. In terms of recall and VMRR, PW-DR performs the best. Specifically, we hold that recall and VMRR are better metrics than precision and F1 to gauge the performance between different strategies. Recall that our goal is to find the most suitable services following a specific service. It means that we care more about how many adjacent services we can fetch, instead of how many candidates are adjacent services. The higher recall means higher possibility to help a user increase the efficiency of service composition. Besides, for the hit index, we hope it could be as low as possible. The higher VMRR could encourage users to reuse the recommended services much more. For example, if a service exists in the ground truth, e.g., $c_{i}$ is followed by two services, say $c_{j}$ and $c_{k}$, which hit at index 1 and 4 of the recommended list, respectively. The precision is only 40\%. However, it is acceptable for the user in real practice to chain one of $c_{j}$ and $c_{k}$ with $c_{i}$, especially in a scenario that there are more than hundreds of services in the whole repository, which makes it tedious and time-consuming to find a suitable service at the next step. Therefore, our experiments have demonstrated that PW-based strategies outperform others.


Table \ref{tab_overall_result} shows that the recall remains around 70\% for all strategies at $K = 5$, meaning that in the ground truth, over half of adjacent services are found if we recommend 5 services.  
If service $c_{i}$ has four adjacent services, about three of them will be returned to the user with recommendation size of 5. Especially, the VMRR@5 is around 0.90. It means that generally the first or the second entry in the recommended list hits the ground truth. We thus have demonstrated that our approach is quite useful to help users save time to composite a workflow by reusing recommended services.

\emph{RQ2: Should duplicates from the generated sequences be removed?} \cite{b36} mentioned that different NLP models are affected by duplicates in different ways. So, it is a good practice to know about the effect of duplicates. Table \ref{tab_overall_result} shows that, for all generation strategies, removing duplicates achieves better performance than remaining them in the training set. 
As for PW-based strategies, due to the fact that the higher probability of a linkage is, the more potential it would be traversed while walking, which makes it more likely to generate duplicate sequences, PW-DR is better than PW in terms of all metrics. 
Therefore, we suggest to investigate the impact of duplicate sequences in advance, when applying language modeling techniques to learn service representations.

\section{Conclusions \& Future Work}
In this paper, we have applied deep learning methods in NLP to study workflow recommendation problem. Workflow composition process is viewed as a sequential service generation process, and the problem of service recommendation is formalized as a problem of next word prediction in NLP. A knowledge graph is constructed on top of workflow provenance, and we develop three strategies to create a corpus of service sequences from the knowledge graph. Deep learning method in NLP is then applied to learn service representations. The resulted service embeddings are used to support incremental workflow composition at run-time, associated with structural information embedded in the knowledge graph.

In the future, we plan to extend our research in the following three directions. First, we plan to take into account users' profile information to enable personalized workflow recommendation. Second, we plan to seamlessly integrate sequential service invocation dependency with hierarchical graph structure to further enhance recommendation quality. Third, to motivate and encourage more research work on service representation learning, we plan to build an open repository supporting a diverse categories of workflow formats. 

\section*{Acknowledgement}
Our work is partially supported by National Aeronautics and Space Administration under grants 80NSSC21K0576 and 80NSSC21K0253.


\vspace{12pt}

\begin{thebibliography}{00}
\bibitem{b1} A. L. Lemos, F. Daniel, \& B. Benatallah, ``Web Service Composition: A Survey of Techniques and Tools", \emph{ACM Computing Surveys (CSUR)}, 48(3), 2015, pp. 1-41.
\bibitem{b2} W. Tan, J. Zhang, \& I. Foster, ``Network Analysis of Scientific Workflows: A Gateway to Reuse", \emph{Computer}, 43(9), 2010, pp. 54-61.
\bibitem{b3} C. A. Goble, J. Bhagat, S. Aleksejevs, D. Cruickshank, D. Michaelides, D. Newman, M. Borkum, S. Bechhofer, M. Roos, P. Li, \& D. D. Roure, ``myExperiment: A Repository and Social Network for the Sharing of Bioinformatics Workflows", \emph{Nucleic Acids Research}, 38, 2010, pp. W677-W682.
\bibitem{b4} J. Zhang, M. Pourreza, S. Lee, R. Nemani, \& T. J. Lee, ``Unit of Work Supporting Generative Scientific Workflow Recommendation", in Proceedings of International Conference on Service-Oriented Computing, Nov. 2018, pp. 446-462.
\bibitem{b5} J. Zhang, W. Wang, X. Wei, C. Lee, S. Lee, L. Pan, \& T. J. Lee, ``Climate Analytics Workflow Recommendation as A Service-provenance-driven Automatic Workflow Mashup", in Proceedings of IEEE International Conference on Web Services, Jun. 2015, pp. 89-97.
\bibitem{b6} D. Hull, R. Stevens, P. Lord, C. Wroe, \& C. Goble, ``Treating Shimantic Web Syndrome With Ontologies", in Proceedings of The 1st AKT Workshop on Semantic Web Services (AKT-SWS04), Dec. 2004, pp. 1-4.
\bibitem{b7} J. Zhang, W. Tan, J. Alexander, I. Foster, \& R. Madduri, ``Recommend-as-you-go: A Novel Approach Supporting Services-oriented Scientific Workflow Reuse", in Proceedings of IEEE International Conference on Services Computing, Jul. 2011, pp. 48-55.
\bibitem{b8} G. Shani, D. Heckerman, \& R. I. Brafman, ``An MDP-based Recommender System", \emph{Journal of Machine Learning Research}, 6(9), 2005, pp. 1265-1295.
\bibitem{b9} F. Yu, Q. Liu, S. Wu, L. Wang, \& T. Tan, ``A Dynamic Recurrent Model for Next Basket Recommendation", in Proceedings of The 39th ACM SIGIR International Conference on Research and Development in Information Retrieval, Jul. 2016, pp. 729-732.
\bibitem{b10} S. Wang, L. Hu, \& L. Cao, ``Perceiving the Next Choice with Comprehensive Transaction Embeddings for Online Recommendation", in Proceedings of Joint European Conference on Machine Learning and Knowledge Discovery in Databases, Sep. 2017, pp. 285-302.
\bibitem{b11} A. Hard, K. Rao, R. Mathews, F. Beaufays, S. Augenstein, H. Eichner, C. Kiddon, \& D. Ramage, ``Federated Learning for Mobile Keyboard Prediction", arXiv, arXiv:1811.03604, 2018.
\bibitem{b12} Y. Zhang, H. Dai, C. Xu, J. Feng, T. Wang, J. Bian, B. Wang, \& T. Y. Liu, ``Sequential Click Prediction for Sponsored Search with Recurrent Neural Networks", AAAI, 28(1), Jun. 2014, pp. 1369-1376.
\bibitem{b13} S. Rendle, C. Freudenthaler, \& L. Schmidt-Thieme, ``Factorizing Personalized Markov Chains for Next-basket Recommendation", in Proceedings of The 19th International Conference on World Wide Web, Apr. 2010, pp. 811-820.
\bibitem{b14} B. Hidasi, A. Karatzoglou, L. Baltrunas, \& D. Tikk, ``Session-based Recommendations with Recurrent Neural Networks", arXiv preprint, arXiv:1511.06939, 2015.
\bibitem{b15} T. Mikolov, K. Chen, G. Corrado, \& J. Dean, ``Efficient Estimation of Word Representations in Vector Space", arXiv preprint, arXiv:1301.3781, 2013.
\bibitem{b16} I. Paik, W. Chen, \& M. N. Huhns, ``A Scalable Architecture for Automatic Service Composition", \emph{IEEE Transactions on Services Computing}, 7(1), 2012, pp. 82-95.
\bibitem{b17} S. R. Chowdhury, F. Daniel, \& F. Casati, ``Efficient, Interactive Recommendation of Mashup Composition Knowledge", in Proceedings of International Conference on Service-Oriented Computing, Dec. 2011, pp. 374-388.
\bibitem{b18} O. Greenshpan, T. Milo, \& N. Polyzotis, ``Autocompletion for Mashups", in Proceedings of the VLDB Endowment, 2(1), 2009, pp. 538-549.
\bibitem{b19} H. Elmeleegy, A. Ivan, R. Akkiraju, \& R. Goodwin, ``Mashup Advisor: A Recommendation Tool for Mashup Development", in Proceedings of IEEE International Conference on Web Services, Sep. 2008, pp. 337-344.
\bibitem{b20} B. Xia, Y. Fan, W. Tan, K. Huang, J. Zhang, \& C. Wu, ``Category-aware API Clustering and Distributed Recommendation for Automatic Mashup Creation", \emph{IEEE Transactions on Services Computing}, 8(5), 2014, pp. 674-687.
\bibitem{b21} D. Koop, C. E. Scheidegger, S. P. Callahan, J. Freire, \& C. T. Silva, ``Viscomplete: Automating Suggestions for Visualization Pipelines", \emph{IEEE Transactions on Visualization and Computer Graphics}, 14(6), 2008, pp. 1691-1698.
\bibitem{b22} S. Deng, D. Wang, Y. Li, B. Cao, J. Yin, Z. Wu, \& M. Zhou, ``A Recommendation System to Facilitate Business Process Modeling", \emph{IEEE Transactions on Cybernetics}, 47(6), 2016, pp. 1380-1394.
\bibitem{b23} J. Zhang, Q. Liu, \& K. Xu, ``FlowRecommender: A Workflow Recommendation Technique for Process Provenance", in Proceedings of The 8th Australasian Data Mining Conference, 2009, pp. 55-61.
\bibitem{b24} S. Smirnov, M. Weidlich, J. Mendling, \& M. Weske, ``Action Patterns in Business Process Models", in Proceedings of The 7th International Conference on Service-Oriented computing, 2009, pp. 115-129.
\bibitem{b25} Y. Bengio, A. Courville, \& P. Vincent, ``Representation Learning: A Review and New Perspectives", \emph{IEEE Transactions on Pattern Analysis and Machine Intelligence}, 35(8), 2013, pp. 1798-1828.
\bibitem{b26} G. E. Hinton, L. Deng, D. Yu, G. E. Dahl, A. Mohamed, N. Jaitly, A. W. Senior, V. Vanhoucke, P. Nguyen, T. N. Sainath, \& B. Kingsbury, ``Deep Neural Networks for Acoustic Modeling in Speech Recognition: The Shared Views of Four Research Groups", \emph{IEEE Signal Processing Magazine}, 29(6), 2012, pp. 82-97.
\bibitem{b27} N. Boulanger-Lewandowski, Y. Bengio, \& P. Vincent, ``Modeling Temporal Dependencies in High-Dimensional Sequences: Application to Polyphonic Music Generation and Transcription", in Proceedings of International Conference on Machine Learning, 2012, pp. 1159-1166.
\bibitem{b28} A. Bordes, X. Glorot, J. Weston, \& Y. Bengio, ``Joint Learning of Words and Meaning Representations for Open-Text Semantic Parsing", in Proceedings of The 15th International Conference on Artificial Intelligence and Statistics, Mar. 2012, pp. 127-135.
\bibitem{b29} H. Paulheim, V. Tresp, \& Z. Liu, ``Representation Learning for the Semantic Web", \emph{Journal of Web Semantics}, 2020, pp. 61-62, 100570.
\bibitem{b30} C. Li, R. Zhang, J. Huai, X. Guo, \& H. Sun, ``A Probabilistic Approach for Web Service Discovery", in Proceedings of IEEE International Conference on Services Computing, Jun. 2013, pp. 49-56.
\bibitem{b31} D. M. Blei, A. Y. Ng, \& M. I. Jordan, ``Latent Dirichlet Allocation", \emph{Journal of Machine Learning Research}, vol. 3, 2003, pp. 993-1022.
\bibitem{b32} Y. Zhong, Y. Fan, W. Tan, \& J. Zhang, ``Web Service Recommendation with Reconstructed Profile from Mashup Descriptions", \emph{IEEE Transactions on Automation Science and Engineering}, 15(2), 2016, pp. 468-478.
\bibitem{b33} M. Rosen-Zvi, T. Griffiths, M. Steyvers, \& P. Smyth, ``The Author-Topic Model for Authors and Documents", in Proceedings of Conference on Uncertainty in Artificial Intelligence, 2004, pp. 487-494.
\bibitem{b34} J. Zhang, Y. Fan, J. Zhang, \& B. Bai, ``Learning to Build Accurate Service Representations and Visualization", \emph{IEEE Transactions on Services Computing}, Jun. 2020.
\bibitem{b35} A. Mnih, \& G. E. Hinton, ``A Scalable Hierarchical Distributed Language Model", in Proceedings of Conference on Neural Information Processing Systems, 2009, pp. 1081-1088.
\bibitem{b36} A. Schofield, L. Thompson, \& D. Mimno, ``Quantifying the Effects of Text Duplication on Semantic Models", in Proceedings of the 2017 Conference on Empirical Methods in Natural Language Processing, 2017, pp. 2737-2747.
\bibitem{b37} B. Perozzi, R. Al-Rfou, \& S. Skiena, ``Deepwalk: Online Learning of Social Representations", in Proceedings of The 20th ACM International Conference on Knowledge Discovery and Data Mining, Aug. 2014, pp. 701-710.
\bibitem{b38} X. He, L. Liao, H. Zhang, L. Nie, X. Hu, \& T. S. Chua, ``Neural Collaborative Filtering", in Proceedings of the 26th International Conference on World Wide Web, Apr. 2017, pp. 173-182.
\bibitem{b39} W. C. Kang, \& J. McAuley, ``Self-Attentive Sequential Recommendation", in Proceedings of IEEE International Conference on Data Mining, Nov. 2018, pp. 197-206.
\bibitem{b40} J. Tang, \& K. Wang, ``Personalized Top-N Sequential Recommendation via Convolutional Sequence Embedding", in Proceedings of The 11th ACM International Conference on Web Search and Data Mining, Feb. 2018, pp. 565-573.
\bibitem{b41} Y. Zhang, M. Zhang, X. Zheng, \& D. E. Perry, ``Service2vec: A Vector Representation for Web Services", in Proceddings of IEEE International Conference on Web Services, Jun. 2017, pp. 890-893.
\bibitem{b42} S. Wang, Z. Wang, \& X. Xu, ``Mining Bilateral Patterns as Priori Knowledge for Efficient Service Composition", in Proceedings of IEEE International Conference on Web Services, Jun. 2016, pp. 65-72.
\bibitem{b43} K. Stoitsas, ``The Use of Word Embeddings for Cyberbullying Detection in Social Media", Doctoral dissertation, Tilburg University, Jul. 2018.
\end{thebibliography}
\end{document}